\pdfoutput=1
\documentclass[11pt]{article}
\usepackage[T1]{fontenc}
\usepackage{lmodern}
\usepackage[utf8]{inputenc}
\usepackage[margin=1in]{geometry}
\usepackage{amsmath,amssymb}
\usepackage{graphicx}
\usepackage{xcolor}
\usepackage{booktabs}
\usepackage{placeins}
\usepackage[round,authoryear]{natbib}
\usepackage{authblk}
\usepackage{xurl}
\usepackage[colorlinks=true,linkcolor=blue!50!black,citecolor=blue!50!black,urlcolor=blue!50!black]{hyperref}
\usepackage[font=small,labelfont=bf]{caption}
\graphicspath{{figures/}}

\title{State-resolved multimodal contributions to stratospheric polar vortex predictability}
\author[1,2,$\dagger$]{Shuo Yang}
\author[1,2,$\dagger$]{Dan Zhao}
\author[1,2,$*$]{Yongwen Zhang}
\author[1,2]{Tingting Xue}
\author[1,2,$*$]{Chunhua Zeng}
\author[3,$*$]{Xiaosong Chen}
\affil[1]{Yunnan Key Laboratory of Complex Systems and Brain-Inspired Intelligence, Kunming University of Science and Technology, Kunming 650500, Yunnan, China}
\affil[2]{Faculty of Science, Kunming University of Science and Technology, Kunming 650500, Yunnan, China}
\affil[3]{Institute for Advanced Study in Physics and School of Physics, Zhejiang University, Hangzhou 310058, China}
\affil[$\dagger$]{These authors contributed equally to this work.}
\affil[$*$]{Corresponding authors: Yongwen Zhang (\href{mailto:zhangyongwen77@gmail.com}{zhangyongwen77@gmail.com}), Chunhua Zeng (\href{mailto:zchh2009@126.com}{zchh2009@126.com}), Xiaosong Chen (\href{mailto:chenxs@zju.edu.cn}{chenxs@zju.edu.cn})}
\date{}

\begin{document}
\maketitle

\begin{abstract}
The dynamical basis of stratospheric polar vortex (SPV) predictability remains unclear, particularly the relative roles of persistence, structural variability, and cross-level coupling. Here we combine eigen microstate theory with mesoscopic Granger-causality analysis of ERA5 geopotential height fields to quantify the predictive contributions of eigen microstates. We show that short-term predictability is dominated by the persistence of the leading stratospheric state, whereas higher-order stratospheric structures and tropospheric variability provide additional predictive information at subseasonal leads. Including these additional states increases the all-winter 30-day prediction correlation from 0.24 to 0.33 and reduces forecast error. These contributions exhibit strong lead-time dependence and vary between winters with and without sudden stratospheric warmings. Part of this predictive information overlaps with that contained in preceding 100 hPa eddy heat flux, linking state-resolved predictability to wave-related circulation variability. Our results unify SPV predictability within a multimodal, state-resolved framework and provide a physically interpretable pathway.
\end{abstract}

\section{Introduction}
The stratospheric polar vortex (SPV) is a major source of subseasonal-to-seasonal predictability in the Northern Hemisphere, exerting a strong influence on midlatitude weather through downward coupling to the troposphere \citep{Baldwin2001,Domeisen2020b}. This downward influence modulates large-scale circulation patterns such as the Arctic Oscillation and Northern Annular Mode \citep{Baldwin1999,Thompson2000,Thompson2000b}, with important impacts on surface climate variability and extreme events \citep{Kidston2015,Domeisen2020b}.

Despite substantial progress in characterizing its variability and impacts, a fundamental question remains unresolved: what determines the dynamical origin of SPV predictability? While enhanced forecast skill is often linked to stratospheric circulation states \citep{Scaife2014,Tripathi2015,Richter2020}, the extratropical stratosphere also exhibits extended predictability relative to the troposphere \citep{Domeisen2020a}. It therefore remains unclear whether SPV predictability arises primarily from this bulk memory or from more complex interactions among multiple circulation structures.

Previous studies have identified key dynamical processes governing SPV variability, including planetary wave propagation and wave--mean-flow interaction, which control both the strength and structure of the vortex \citep{Charney1961,Holton1976,Matsuno1971}. In particular, strong upward-propagating Rossby waves from the troposphere can perturb and even disrupt the vortex, leading to sudden stratospheric warming (SSW) events \citep{Matsuno1971,Butler2015,Baldwin2021}. The response also depends on the stratospheric background state: nonlinear self-tuning resonance can contribute to vortex disruption in idealized descriptions of splitting events \citep{Baldwin2021,MatthewmanEsler2011}. In addition, stratosphere--troposphere coupling has been recognized as a major pathway through which stratospheric variability influences surface climate and enhances subseasonal predictability \citep{Kidston2015,Butler2018,Domeisen2020b}.

From a statistical perspective, dominant variability is often characterized using EOF- or PCA-based methods \citep{Hannachi2007,Jolliffe2002,Monahan2009}, and predictive relationships are explored using regression-based or data-driven approaches \citep{Kretschmer2017,WuZ2022}. However, a key question is how multiple circulation states jointly contribute to predictability across timescales and dynamical regimes. In particular, it remains unclear whether extended-range predictability emerges primarily from a single dominant mode or from the collective contribution of multiple interacting states, and how this balance may shift between normal winters and winters with SSWs, given the distinct dynamical evolution and predictability of extreme vortex events \citep{Gray2020,WuRW2022}.

Here we propose that SPV predictability has a multimodal origin, arising from the joint evolution of multiple stratospheric and tropospheric states. To test this hypothesis, we adopt Eigen Microstate Theory (EMT), first introduced by \citet{Hu2019} and \citet{Sun2021}, and later extended to nonequilibrium systems and climate prediction problems, including the increased predictability of extreme El Ni\~no from decadal interbasin interaction \citep{Zhang2024,Ma2024}. EMT shares its SVD representation with covariance-based EOF/PCA for the same anomaly matrix and weighting; its statistical-physics interpretation treats the sampled configurations as an ensemble described by eigen microstates and their ensemble weights. Orthogonality does not imply that these states evolve independently. We further combine EMT with a Granger-causality framework, in which additional predictive information is assessed at the level of eigen microstates conditional on the history of the leading stratospheric state \citep{Granger1969,Runge2019}. This mesoscopic representation captures circulation structures and enables us to quantify how distinct circulation states contribute to the predictability of the system. Using ERA5 data (1948--2023), we assess how vortex memory and multimodal information contribute to SPV predictability across forecast leads and winter regimes.

\section{Data and Methods}\label{sec:methods}

\subsection{Data and preprocessing}\label{sec:data}
We analyze geopotential height fields from the ERA5 reanalysis at 10~hPa and 500~hPa, representing the stratospheric and tropospheric circulation, respectively \citep{Hersbach2020}. The analysis domain is the Arctic polar cap ($65^\circ$--$90^\circ$N) on a $2.5^\circ\times2.5^\circ$ grid, and the study period spans 1948--2023. Daily mean fields are computed from 6-hourly data. To examine subseasonal variability, a multi-year daily climatology is removed at each grid point, yielding deseasonalized anomaly fields. We denote the anomaly fields at time $t$ as $\mathbf{x}^{S}(t)$ and $\mathbf{x}^{T}(t)$, representing the stratospheric (10~hPa) and tropospheric (500~hPa) circulation, respectively. The sample contains 75 winters from 1948/49 to 2022/23.

The SPV index is defined as the normalized polar-cap-averaged geopotential height anomaly at 10~hPa, with monthly linear trends removed prior to normalization \citep{Shan2022}. The polar-cap average uses cosine-latitude weights. For SPV, an intercept and linear trend in continuous time are fitted separately to the daily anomalies in each calendar month, and the residuals are centered and divided by a common November--April standard deviation. Positive SPV values denote positive polar-cap height anomalies. Winters are classified as SSW winters or normal winters based on the reversal of zonal-mean zonal wind at 10~hPa and $60^\circ$N \citep{Charlton2007}. The event catalogue contains 47 central dates in 39 SSW winters and is listed in Text~S3; the remaining 36 winters are normal winters. These observed whole-winter classes define the class-conditional analyses below.

We focus on SPV variability during the extended winter season from November to April. To construct predictors with sufficient lead time, data from September to April are additionally used. The 60-day histories required at the earliest forecast origins also use July--August fields projected onto the same winter spatial basis. Climatologies, trends, normalization, and spatial bases are estimated within each training sample for forecast validation, as described in Section~\ref{sec:prediction}.

\subsection{Eigen-microstate decomposition}\label{sec:emt}
To extract dominant circulation structures, we apply EMT separately to the stratosphere and troposphere.

For each layer, the EMT decomposition is performed over the concatenated seasonal segments from September of each year to April of the following year across all winters in the estimation sample. The available estimation dates are September~2--April~30, excluding February~29; grid points have equal weight in the decomposition.

For the stratosphere, anomaly fields are assembled into an ensemble matrix \citep{Hu2019,Sun2021}:
\begin{equation}
\mathbf{Z}^{S}=\big[\mathbf{x}^{S}(t_1)-\bar{\mathbf{x}}^{S},\ldots,\mathbf{x}^{S}(t_M)-\bar{\mathbf{x}}^{S}\big],
\label{eq:emt_ensemble}
\end{equation}
where
\begin{equation}
\bar{\mathbf{x}}^{S}=\frac{1}{M}\sum_{m=1}^{M}\mathbf{x}^{S}(t_m)
\label{eq:emt_mean}
\end{equation}
is the time mean over the full sampled period.

We normalize the ensemble matrix using the Frobenius norm,
\begin{equation}
\widetilde{\mathbf{Z}}^{S}=\frac{\mathbf{Z}^{S}}{\|\mathbf{Z}^{S}\|_F},
\label{eq:emt_normalization}
\end{equation}
where $\|\cdot\|_F$ denotes the Frobenius norm.

Singular value decomposition (SVD) is then applied:
\begin{equation}
\widetilde{\mathbf{Z}}^{S}=\mathbf{U}^{S}\mathbf{\Sigma}^{S}(\mathbf{V}^{S})^{\top}.
\label{eq:emt_svd}
\end{equation}
For the stratosphere, we denote the $i$th eigen microstate by $S_i$, whose temporal coefficient $a_i^{S}(t)$ is given by the $(t,i)$ element of $\mathbf{V}^{S}$, and whose spatial pattern is given by the $i$th column of $\mathbf{U}^{S}$. Similarly, for the troposphere, we obtain eigen microstates $T_i$ with temporal coefficients $a_i^{T}(t)$, where the spatial patterns are given by the columns of $\mathbf{U}^{T}$. The contribution of each state is quantified by
\begin{equation}
p_i^{X}=(\sigma_i^{X})^2,\qquad X\in\{S,T\},
\label{eq:emt_weight}
\end{equation}
which represents the fraction of variance explained by each eigen microstate.

Here $\sigma_i^{X}$ is the $i$th singular value of the normalized ensemble matrix, and $\sum_i p_i^{X}=1$ for the complete decomposition. Note that lagged and nonlinear dependencies between the temporal coefficients can represent coupled circulation variability, and several eigen microstates may jointly describe one vortex disturbance.

For descriptive spatial analyses, all 75 winters define a common reference basis. For forecast validation, each basis uses only the training winters. Coefficients on held-out dates and on July--August history dates are obtained by projection onto that basis. The projection and conversion to the temporal $\mathbf V$ coordinates are described in Text~S1.

\subsection{Identification of SPV-related states}\label{sec:laggedcorr}
We next identify which eigen microstates are most relevant to SPV variability.

For each state, we compute the lagged correlation between its temporal coefficient and the SPV index:
\begin{equation}
r_i^{X}(\tau)=\operatorname{corr}\big(a_i^{X}(t-\tau),\mathrm{SPV}(t)\big),
\label{eq:lagged_correlation}
\end{equation}
where $\tau\geq0$ denotes the lag, and $t$ is restricted to the period from November to April.

For each state, we identify the lag that maximizes the absolute correlation for $\tau\geq0$:
\begin{equation}
\tau_i^*=\underset{\tau\in[0,\tau_{\max}]}{\operatorname{arg\,max}}\,|r_i^{X}(\tau)|,
\label{eq:maximum_lag}
\end{equation}
and define the predictive relevance as
\begin{equation}
R_i^{X}=r_i^{X}(\tau_i^*),
\label{eq:lag_relevance}
\end{equation}
where $\tau_{\max}=60$ days. Lagged correlations are calculated at integer lags of 0--60 calendar days, without pairing data from different winters.

Statistical significance is assessed using phase-randomized surrogate data. We generate 999 surrogate SPV series within continuous winter segments and repeat the same maximum-absolute-correlation search for each realization. The 2.5th and 97.5th percentiles of the signed surrogate extrema provide the reference range for Figures~\ref{fig:statesS}a and \ref{fig:statesT}a. Pointwise lag curves are shown in Figures~S1 and S2; the surrogate construction is described in Text~S2. Guided by these descriptive associations and the retention of leading variance-bearing structures, we retain six stratospheric and seven tropospheric eigen microstates. These dimensions are fixed throughout forecast validation.

\subsection{Prediction framework}\label{sec:prediction}
We define the prediction target as
{
\begin{equation}
z(t)=\mathrm{SPV}(t).
\label{eq:prediction_target}
\end{equation}
}
Because the leading stratospheric state is highly correlated with the SPV index ($r=0.979$ at zero lag; Figure~S1a), we use its 60-day history as a baseline representing vortex memory.

Baseline model: for initialization date $t$, lead $h$, and winter group $g$ (All, SSW, or normal),
{
\begin{equation}
\widehat z^{(B)}_{g,h}(t)=c^{(B)}_{g,h}
+\sum_{\ell=0}^{59}\phi^{(B)}_{g,h,\ell}a_1^{S}(t-\ell).
\label{eq:baseline_spv}
\end{equation}
}
Multimodal model:

Let $\mathcal S$ and $\mathcal T$ denote the sets of selected stratospheric and tropospheric states, respectively. Since the leading stratospheric state is already included in the history block, it is excluded from the additional stratospheric predictor set. The multimodal model is written as
{
\begin{equation}
\widehat z^{(M)}_{g,h}(t)=c^{(M)}_{g,h}
+\sum_{\ell=0}^{59}\phi^{(M)}_{g,h,\ell}a_1^{S}(t-\ell)
+\sum_{i\in\mathcal S}b_{g,h,i}a_i^{S}(t)
+\sum_{j\in\mathcal T}d_{g,h,j}a_j^{T}(t),
\label{eq:direct_spv}
\end{equation}
}
where $\mathcal S=\{2,\ldots,6\}$ and $\mathcal T=\{1,\ldots,7\}$, and $\widehat z_{g,h}(t)$ predicts $z(t+h)$. The history comprises 60 daily values ending at initialization; the other 12 coefficients enter only at initialization. For each model configuration and lead $h=1,\ldots,60$ calendar days, its intercept and all regression coefficients are jointly estimated by unpenalized ordinary least squares (OLS). The baseline and full model contain 61 and 73 parameters, respectively. Two additional configurations add only $S_2$--$S_6$ or only $T_1$--$T_7$ to the same history. The linear equations extract predictive relationships from historical and spatial states generated by nonlinear atmospheric evolution.

Forecast evaluation leaves out one complete winter at a time. In each fold, the other 74 winters determine gridpoint daily climatologies, EMT anomaly means and spatial bases, and the SPV climatology, monthly trends, and normalization. Fields from the held-out winter are projected onto these training-derived bases. Calendar-day climatologies for EMT use the available July--April training fields; spatial basis estimation remains restricted to September--April. The All regression uses the other 74 winters. For a held-out SSW or normal winter, the corresponding class-specific regression uses the other 38 SSW or 35 normal winters, respectively. All three samples share the fold's All-winter preprocessing and spatial basis.

Dates are paired by actual calendar-day subtraction. A valid pair requires the target SPV, all 60 consecutive daily $S_1$ coefficients ending at initialization, and all 12 additional initialization-day coefficients to be available within the same July--June winter. Every model configuration uses these same training and verification pairs at a given lead. The earliest required history date is July~5 for a November~1 target at a 60-day lead; summer fields are projected onto the training winter basis. Calendar details are provided in Text S3.

\subsection{Predictability and Granger causality}\label{sec:granger}\label{sec:conditional_coefficients}
Predictability at lead time $h$ is quantified by the Pearson correlation:
{
\begin{equation}
\mathrm{Skill}_{g}(h)=\operatorname{corr}\big[\widehat z_{g,h}(t),z_{\mathrm{obs}}(t+h)\big],
\label{eq:prediction_skill}
\end{equation}
}
where $\widehat z_{g,h}(t)$ denotes the $h$-day-ahead prediction initialized at time $t$. Predictability is evaluated separately for all winters, SSW winters, and normal winters by pooling their corresponding held-out daily forecast pairs. We also calculate the mean-squared-error ratio
{
\begin{equation}
Q_{g,h}=\frac{\sum_{t\in\mathcal I_{g,h}}[\widehat z_{g,h}(t)-z_{\mathrm{obs}}(t+h)]^2}
{\sum_{t\in\mathcal I_{g,h}}z_{\mathrm{obs}}(t+h)^2},
\label{eq:mse_ratio}
\end{equation}
}
where $\mathcal I_{g,h}$ contains the valid held-out forecast pairs. Thus $Q<1$ indicates lower MSE than a zero-SPV-anomaly forecast on the same samples. Pointwise 95\% percentile intervals are obtained from 2,000 paired whole-winter bootstrap draws, recalculating pooled scores for each draw while keeping the fitted forecasts fixed.

To quantify the contribution of each selected eigen microstate to predictability, we construct a Granger-causal framework \citep{Granger1969} based on the multimodal model in Eq.~\ref{eq:direct_spv}. Specifically, for each additional eigen microstate, we compare the full model with a reduced model in which the corresponding state is excluded. Both models retain all 60 $S_1$ history terms, and the reduced model retains the other 11 current predictors. The nested restriction is $H_0:\beta_j(h)=0$, where $\beta_j$ denotes the relevant $b_i$ or $d_j$ coefficient.

For coefficient estimation and testing, Equation~\ref{eq:direct_spv} is fitted separately to the complete All, SSW, and normal samples (75, 39, and 36 winters), with a common All-winter EMT basis and SPV preprocessing. These descriptive fits express predictors in the unit-length temporal $\mathbf V$ coordinates (Text~S1). We report the absolute values of the 12 additional regression coefficients, in units of standardized SPV per $\mathbf V$-coordinate unit. The single-coefficient test uses a finite-sample-corrected covariance estimate clustered by winter (CR1; \citealp{CameronMiller2015}):
{
\begin{equation}
W_j(h)=\frac{\widehat\beta_j(h)^2}
{\widehat{\operatorname{Var}}_{\mathrm{CR1}}[\widehat\beta_j(h)]},
\qquad W_j(h)\ \dot\sim\ F_{1,G_g-1},
\label{eq:conditional_wald}
\end{equation}
}
with $G_g=75$, 39, or 36 winters. This is equivalent to a two-sided $t_{G_g-1}$ test and addresses the same single zero-coefficient restriction as a classical nested-model RSS test, while allowing within-winter residual dependence through the CR1 variance estimate (Text~S3). A pointwise $p\leq0.05$ indicates a statistically significant conditional contribution, without adjustment for multiple comparisons. In this Granger-causality framework, the contribution concerns information about subsequent SPV conditional on the specified $S_1$ history and the other eigen-microstate predictors. Held-out forecast comparisons separately assess whether additional predictors improve prediction error.

\subsection{Heat-flux information and eigen-microstate overlap}\label{sec:heatflux}

To connect the circulation coordinates with wave activity entering the stratosphere \citep{PolvaniWaugh2004,BirnerAlbers2017}, we calculate 100-hPa meridional eddy heat flux from ERA5 meridional wind and temperature on a $1^\circ$ grid. At 00, 06, 12, and 18~UTC, zonal means are removed from both fields, their product is averaged over longitude, and the result is averaged over $45^\circ$--$75^\circ$N with cosine-latitude weights. The four instantaneous fluxes are then averaged to form daily $H(t)$. This order retains subdaily eddy covariance. Calendar-day climatology estimated from the other 74 winters gives the daily anomaly $H'(t)$. The predictor is
{
\begin{equation}
H_{40}(t)=\frac1{40}\sum_{\ell=0}^{39}H'(t-\ell),\label{eq:h40}
\end{equation}
}
requiring all 40 calendar days and ending at initialization. This fixed multiweek averaging window follows the accumulated-wave-activity perspective of \citet{PolvaniWaugh2004}.

Models $BH$ and $MH$ add $H_{40}(t)$ to $B$ and $M$, respectively, with all parameters refitted within each training fold. On identical held-out pairs, $Q(B)-Q(BH)$ measures the H40 increment beyond the history baseline, $Q(B)-Q(M)$ the increment from the additional eigen microstates, and $Q(M)-Q(MH)$ the remaining H40 increment after those predictors are included. Positive values indicate lower prediction error. Their intervals use the same paired winter-bootstrap procedure.

To locate overlapping prediction information, let $M_{-j}$ denote the full model with predictor $j$ removed, and $M_{-j}H$ the corresponding model with $H_{40}$ added. All remaining parameters are refitted within each training fold. We calculate
{
\begin{equation}
C_j(h)=\big[Q_h(M_{-j})-Q_h(M_{-j}H)\big]
-\big[Q_h(M)-Q_h(MH)\big].\label{eq:heat_overlap}
\end{equation}
}
A positive $C_j$ means that H40 provides a larger increment when predictor $j$ is absent, consistent with overlapping information conditional on the retained predictors. Negative values indicate the opposite change. All comparisons retain the full $S_1$ history. The heatmap displays these signed changes in normalized MSE, providing a predictive-information connection with the heat-flux diagnostic.

\section{Results}
\subsection{Eigen microstates associated with SPV variability}
We first apply EMT to the stratosphere and troposphere using all winters from 1948 to 2023 (see Section~\ref{sec:emt}), focusing on the polar-cap domain ($65^\circ$--$90^\circ$N). Using the lagged-correlation framework defined in Equation~\ref{eq:lagged_correlation}, we quantify the relationship between each eigen microstate and the SPV index.

\begin{figure}[tbp]
\centering
\includegraphics[width=\textwidth]{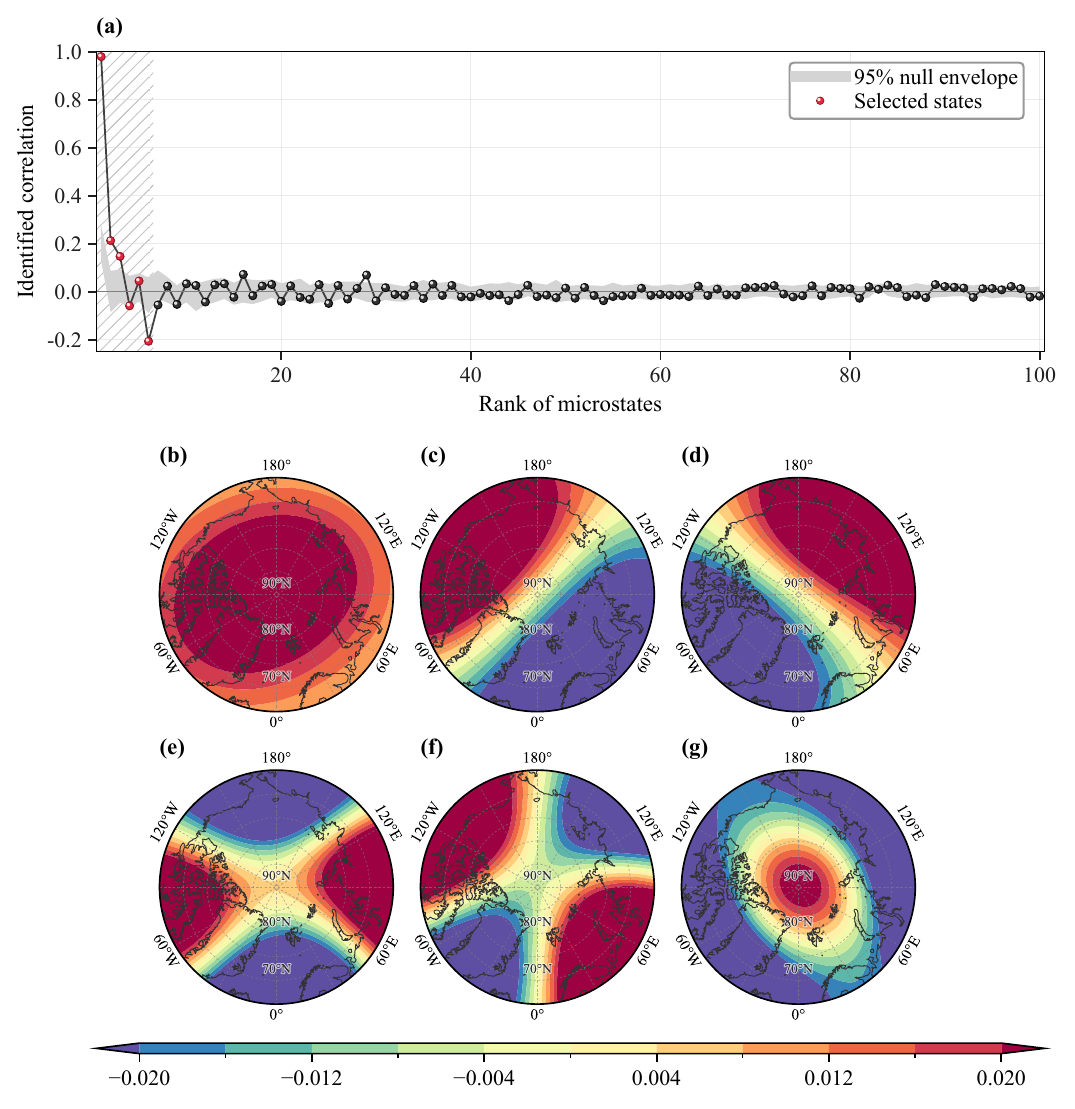}
\caption{Predictive relevance and spatial patterns of the selected stratospheric eigen microstates. (a) Correlation at the maximum absolute correlation over 0--60-day lags for each of the first 100 temporal coefficients. Gray shading gives the 95\% null envelope from 999 phase-randomized SPV surrogates, each repeating the lag search. Red markers and hatching identify the six retained eigen microstates. (b--g) Unit-norm spatial patterns of $S_1$--$S_6$; percentages give fractions of total anomaly variance. The common descriptive basis uses September~2--April~30 anomalies over all 75 winters. Spatial and temporal signs of $S_1$ and $S_2$ are chosen so their correlations in (a) are positive. Colorbar extensions denote values outside the displayed range. The eigen microstates $S_1$--$S_6$ shown in panels (b--g) explain 71.23\%, 13.54\%, 9.22\%, 1.82\%, 1.59\%, and 1.49\% of the total 10-hPa geopotential height anomaly variance, respectively.}
\label{fig:statesS}
\end{figure}

To systematically identify eigen microstates with predictive relevance, we compute $R_i^S$ (Equation~\ref{eq:lag_relevance}) for the leading 100 stratospheric eigen microstates. Figure~\ref{fig:statesS}a shows $R_i^S$ together with the 95\% surrogate reference range. The leading states $S_1$, $S_2$, $S_3$, and $S_6$ lie outside this range. Although $S_4$ and $S_5$ do not, they possess relatively large eigenvalues and are therefore retained to preserve the dominant variance-contributing structures. As a result, six leading stratospheric eigen microstates ($S_1$--$S_6$) are selected for subsequent analysis. These dimensions are fixed throughout forecast validation.

Figures~\ref{fig:statesS}b--g show the spatial patterns of these states over the polar cap. The leading state $S_1$ exhibits a nearly axisymmetric monopole structure (Figure~\ref{fig:statesS}b), representing coherent strengthening and weakening of the vortex consistent with previous studies in which the dominant winter stratospheric variability is primarily associated with changes in polar-vortex intensity \citep{ChenWei2009}. It explains 71.23\% of the polar-cap anomaly variance and corresponds to variability in bulk vortex intensity.

The next pair, $S_2$ and $S_3$, displays dipolar structures (Figures~\ref{fig:statesS}c and \ref{fig:statesS}d) with different longitudinal orientations. These eigen microstates represent large-scale zonal asymmetry within the polar cap. A Fourier decomposition assigns 98.43\% and 98.88\% of their spatial squared amplitudes to zonal wavenumber~1 (Table~S1). The following pair, $S_4$ and $S_5$, exhibits quadrupolar structures (Figures~\ref{fig:statesS}e and \ref{fig:statesS}f), with alternating centers of opposite sign across the polar cap. Their spatial squared amplitudes are dominated by wavenumber~2, accounting for 93.46\% and 90.69\%, respectively. These pairs describe wave-1 and wave-2 components of polar-cap circulation variability. Multiple eigen microstates may jointly describe a vortex disturbance; individual states are not identified one-to-one with displacement or splitting processes.

The higher-order state $S_6$ exhibits a polar-centered structure (Figure~\ref{fig:statesS}g). This state resembles a radially structured higher-order monopole, with an oppositely signed outer region. $S_6$ is predominantly zonally symmetric. Its opposite-signed centers near the pole and over the outer polar cap describe radial variations in height anomalies, complementing the bulk vortex-intensity variations represented by $S_1$.

Supplementary Figure~S1 shows examples of the lagged-correlation functions for representative stratospheric eigen microstates. The leading state $S_1$ is closely related to the SPV index, exhibiting a correlation of $0.979$ at zero lag (Supplementary Figure~S1a), confirming that it captures the bulk variability of the polar vortex. In contrast, higher-order states exhibit asymmetric lagged-correlation structures with extrema at nonzero lags (e.g., $S_2$, with $r=0.213$ at a 21-day lag in Supplementary Figure~S1b), indicating potential predictive relationships with subsequent SPV evolution.

\begin{figure}[tbp]
\centering
\includegraphics[width=\textwidth]{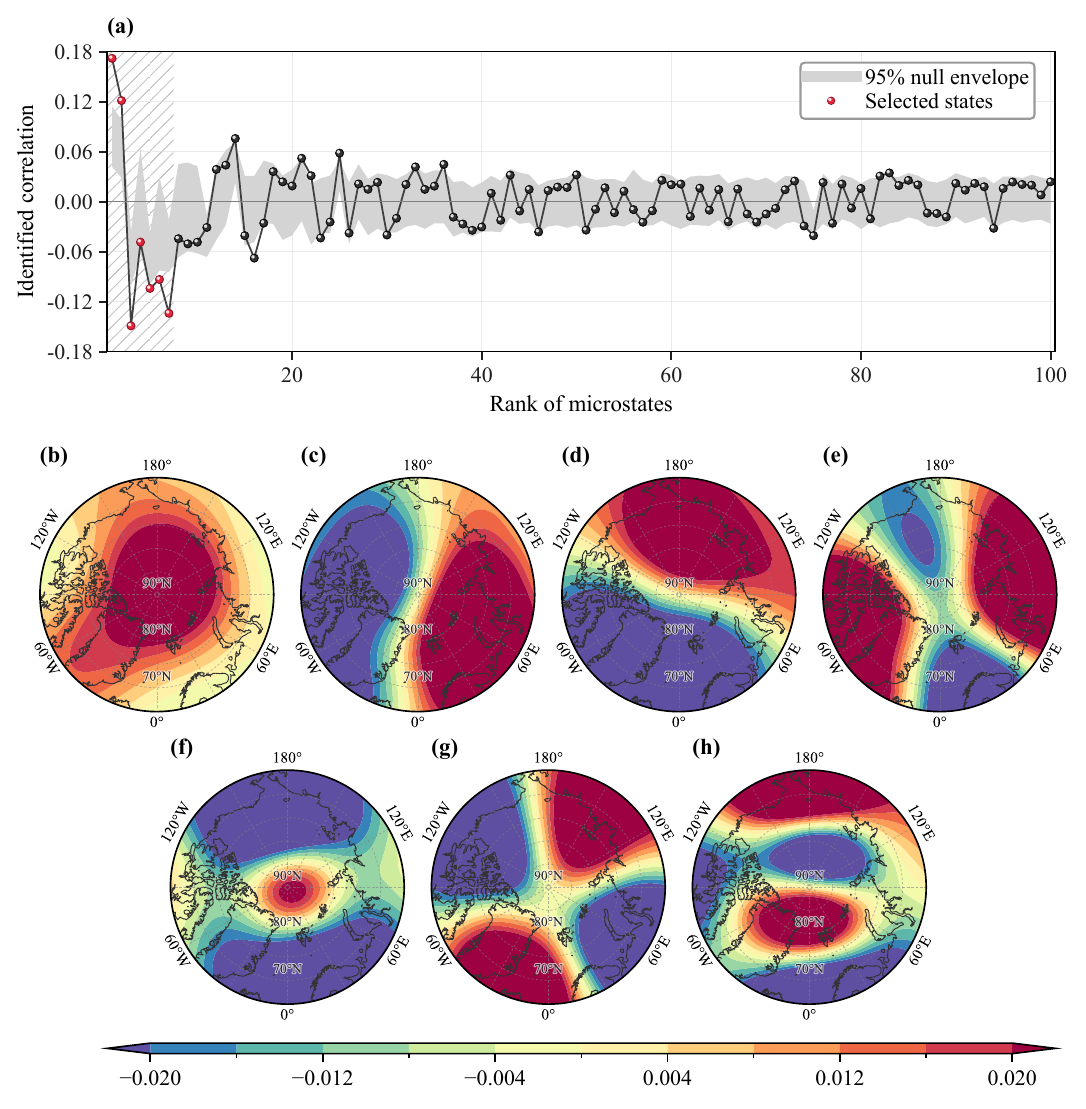}
\caption{Predictive relevance and spatial patterns of the selected tropospheric eigen microstates. (a) Correlations from the same maximum-absolute lag search and surrogate procedure as Figure~\ref{fig:statesS}, using the first 100 temporal coefficients at 500~hPa. Red markers identify the seven retained eigen microstates. (b--h) Unit-norm spatial patterns of $T_1$--$T_7$, with their fractions of total anomaly variance. The spatial basis uses the same 75-winter reference period as Figure~\ref{fig:statesS}. The eigen microstates $T_1$--$T_7$ shown in panels (b--h) explain 34.62\%, 14.11\%, 10.34\%, 7.01\%, 6.69\%, 5.46\%, and 2.81\% of the total 500-hPa geopotential height anomaly variance, respectively.}
\label{fig:statesT}
\end{figure}

A similar analysis for the troposphere identifies seven dominant eigen microstates (Figure~\ref{fig:statesT}a). The lagged-correlation functions of their temporal coefficients (Supplementary Figure~S2) exhibit structured associations with SPV, with extrema at nonzero lags for several states. However, their correlations are generally weaker than those of the stratospheric states.

Compared to the stratospheric states, the tropospheric eigen microstates exhibit stronger spatial asymmetry and more complex multipolar structures (Figures~\ref{fig:statesT}b--h). Such spatial organization is consistent with regional circulation variability associated with large-scale wave activity and teleconnections \citep{Quadrelli2004}. These states provide a complementary description of the lower atmosphere whose predictive contribution can be evaluated jointly with the stratospheric states.

\FloatBarrier

\subsection{Multimodal contributions to SPV predictability}\label{sec:results_prediction}\label{sec:results_skill_variation}

\begin{figure}[tbp]
\centering
\includegraphics[width=\textwidth]{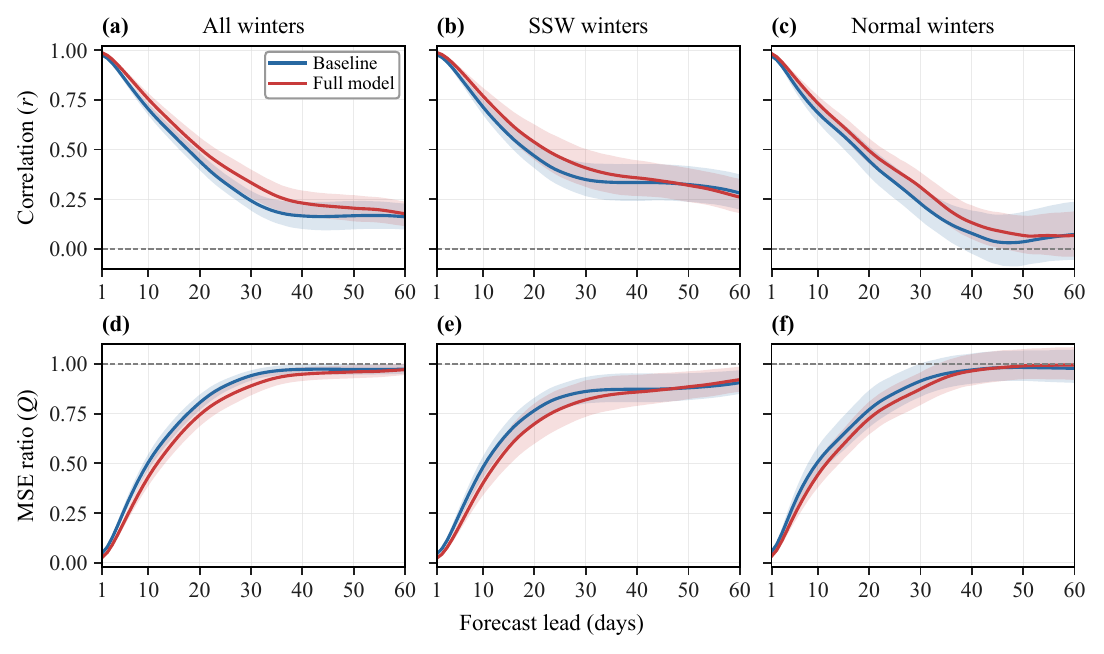}
\caption{Prediction skill measured by correlation (Equation~\ref{eq:prediction_skill}) and the zero-SPV-anomaly MSE ratio $Q$ as a function of lead time for different model configurations and winter regimes. Blue denotes the 60-day $S_1$-history baseline; red denotes the full model adding the origin-day temporal coefficients of $S_2$--$S_6$ and $T_1$--$T_7$. Columns show All, SSW, and normal samples of 75, 39, and 36 held-out winters. (a--c) Prediction--observation correlation; (d--f) forecast MSE relative to a zero-SPV-anomaly forecast, $Q$. Direct regressions use 74, 38, and 35 other winters, respectively, with preprocessing and EMT estimated from the other 74 winters in all groups. Models share valid forecast pairs at each lead. Shading shows pointwise 95\% paired winter-bootstrap intervals with forecasts held fixed.}
\label{fig:corr}
\end{figure}

We next assess how the selected eigen microstates contribute to SPV predictability as a function of lead time. Figure~\ref{fig:corr} compares prediction correlations and mean-squared errors under different model configurations. A restricted baseline model, relying solely on the 60-day history of the leading stratospheric state $S_1$, exhibits a decay in forecast correlation with increasing lead time (Figures~\ref{fig:corr}a--c). At short lead times, the model retains relatively high skill, reflecting the intrinsic persistence of bulk vortex intensity. However, this skill decreases substantially at longer lead times.

Incorporating the full multimodal structure, by including both stratospheric and tropospheric eigen microstates (Equation~\ref{eq:direct_spv}), improves prediction skill over the following several weeks and slows the decay of predictability. For All winters, adding the origin-day coefficients of $S_2$--$S_6$ and $T_1$--$T_7$ to the 60-day $S_1$ history increases prediction--observation correlation from 0.800 to 0.841 at 7~days and from 0.568 to 0.625 at 15~days (Figure~\ref{fig:corr}a). The corresponding zero-anomaly MSE ratio $Q$ decreases from 0.360 to 0.293 and from 0.677 to 0.610 (Figure~\ref{fig:corr}d). At 30~days, correlation increases from 0.242 to 0.333 and $Q$ decreases from 0.941 to 0.890; the paired reduction in $Q$ is 0.051, with a pointwise 95\% winter-bootstrap interval of 0.012--0.085. Thus, the additional coefficients improve both covariation with SPV and squared-error performance beyond information already present in the $S_1$ history.

This behavior depends on the winter group and lead time. At 15~days in SSW winters, correlation increases from 0.574 to 0.638 and $Q$ decreases from 0.657 to 0.581 (Figures~\ref{fig:corr}b and \ref{fig:corr}e). At 30~days, the SSW values change from 0.349 to 0.408 and from 0.862 to 0.820, respectively, whereas the normal values change from 0.228 to 0.311 and from 0.914 to 0.874 (Figures~\ref{fig:corr}c and \ref{fig:corr}f). The 30-day point estimates therefore improve in both classes, although the paired $Q$-gain intervals include zero for each class separately. By 60~days, the All-winter improvement is small ($r=0.161$ versus 0.177; $Q=0.974$ versus 0.971), and the full model has higher $Q$ than the baseline in both class-specific evaluations. At these longer leads, most of the retained SSW-class skill is already represented by the history baseline.

To further disentangle the sources of this improvement, we separately evaluate the contributions from stratospheric and tropospheric states (Figure~S3). At 30~days for All winters, adding only $S_2$--$S_6$ gives $r=0.286$ and $Q=0.919$, while adding only $T_1$--$T_7$ gives $r=0.290$ and $Q=0.916$, compared with $r=0.242$ and $Q=0.941$ for the baseline. The full model attains higher correlation and lower error than either restricted extension at this lead. In All winters, the stratospheric extension yields higher correlation and lower error during the first three weeks, while the two extensions perform similarly around four weeks. The combined configuration thus draws on complementary information at the two levels.

The predictive gain beyond the $S_1$-history baseline also varies with target month, with pronounced improvements in December and January (Figures~S4--S6). At 15~days, December has the largest point improvement in both correlation and $Q$ among the six verification months in each winter group. In All winters, December correlation increases from 0.499 to 0.627 and $Q$ decreases from 0.753 to 0.608. At 30~days, January has the largest increments for All and SSW winters: the All-winter correlation increases from 0.206 to 0.407 and $Q$ decreases from 0.959 to 0.846. Spring skill often contains a substantial contribution from the history baseline. For example, in March at 30~days the SSW baseline has $r=0.593$ and $Q=0.650$, compared with $r=0.565$ and $Q=0.695$ for the full model.

Event alignment reveals how skill changes around vortex disruptions (Figure~S7). At 7- and 15-day leads, additional eigen microstates generally improve SPV forecasts before the observed SSW central date. Prediction--observation correlation decreases near and shortly after that date, and the gain at 30~days is less uniform, with wide uncertainty intervals. For many later targets, much of the skill is already captured by the $S_1$-history baseline, consistent with the importance of bulk vortex memory during the subsequent evolution.

Across rolling 10-winter windows, short-range skill remains relatively stable while longer-lead skill exhibits more pronounced decadal fluctuations (Figure~S8). The curves show no clear long-term trend or consistent abrupt shift around 1979, which marks the transition to more extensive satellite observations in ERA5 \citep{Hersbach2020}.

Forecast skill is similar for the two tested EMT estimation seasons (Figure~S9). Re-estimating each training-fold basis from December--February instead of September--April and refitting the prediction equations changes full-model correlation by less than 0.0025 and $Q$ by less than 0.0013 across all three groups and leads. Regression and verification dates remain the same. The retained representation therefore supplies similar predictive information when its spatial basis is estimated only in midwinter, even though the variance and predictability of the circulation vary through the season.

\FloatBarrier
\subsection{State-resolved Granger-causality analysis}\label{sec:results_contribution}\label{sec:results_heatflux}

\begin{figure}[tbp]
\centering
\includegraphics[width=\textwidth]{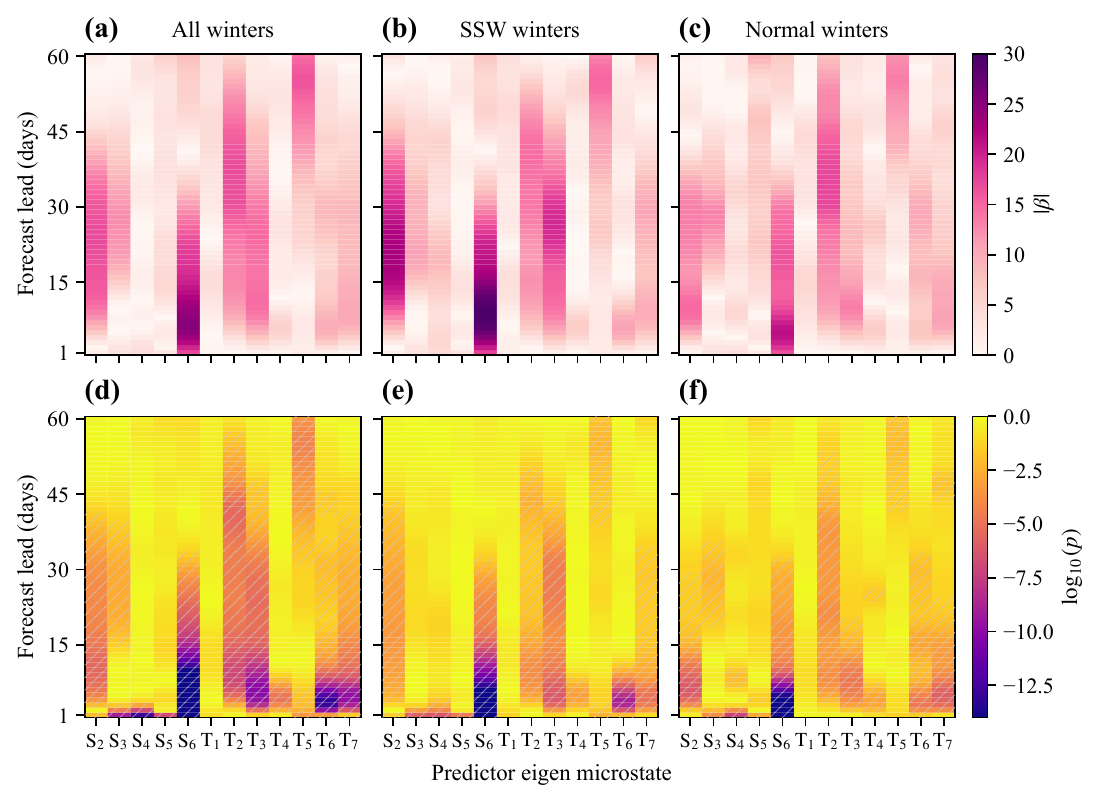}
\caption{Regression coefficients and Granger-causality significance as a function of lead time for individual eigen microstates, conditional on the 60-day $S_1$ history. Columns show descriptive fits to All, SSW, and normal samples in a common 75-winter EMT basis. (a--c) Absolute regression coefficients for the 12 additional eigen-microstate predictors, in SPV standard deviations per unit of the temporal $V$ coordinate. (d--f) $\log_{10}p$ for individual zero-coefficient restrictions, controlling for all history terms and the other 11 current predictors. Tests use winter-cluster CR1 covariance (Equation~\ref{eq:conditional_wald}); white hatching marks pointwise, unadjusted $p\leq0.05$. Values below $\log_{10}p=-14$ share the lowest color.}
\label{fig:sign}
\end{figure}

To further quantify the conditional predictive contributions of individual circulation states, we analyze both the regression coefficients and the Granger-causal significance associated with each additional eigen microstate using the framework defined in Equation~\ref{eq:conditional_wald}. Figure~\ref{fig:sign} summarizes these results for all winters, SSW winters, and normal winters. The 12 coefficients (Figures~\ref{fig:sign}a--c) are estimated jointly with 60~days of $S_1$ history, so their tests concern information beyond that history and the other current coefficients. $S_6$ has a prominent short-lead association in all three winter groups, with its individual coefficient passing the pointwise winter-cluster test through approximately 32~days (Figures~\ref{fig:sign}d--f). $S_2$, $T_2$, and $T_3$ also have sustained conditional associations over intermediate leads, with differences among groups. In particular, the conditional association of $T_3$ extends to about 44~days in SSW winters (Figure~\ref{fig:sign}e) but is concentrated within about 20~days in normal winters (Figure~\ref{fig:sign}f). $T_5$ retains a long-lead conditional association in each group. The contributions of individual eigen microstates thus vary with both forecast lead and winter regime.

\begin{figure}[tbp]
\centering
\includegraphics[width=\textwidth]{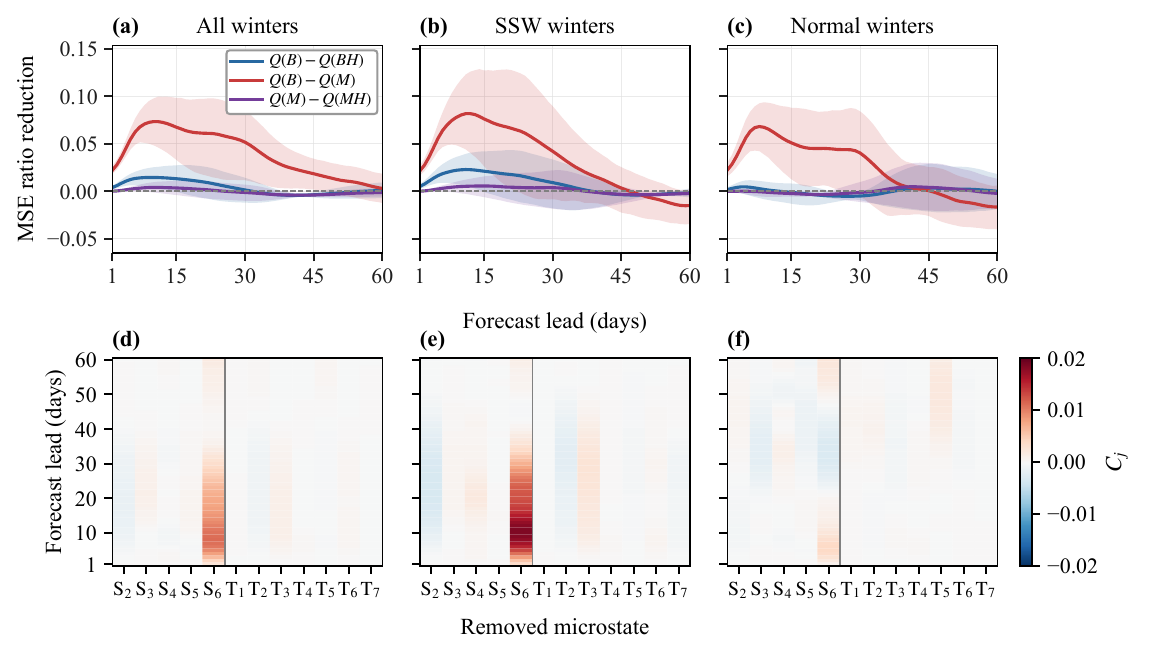}
\caption{Eddy heat flux and eigen-microstate contributions to SPV forecast skill. Columns show All, SSW, and normal held-out samples. (a--c) Reduction in $Q$ from adding $H_{40}$ to the history baseline (blue), adding the 12 eigen-microstate predictors to that baseline (red), and adding $H_{40}$ to the full model (purple). Positive values indicate lower forecast error. Shading shows pointwise 95\% paired winter-bootstrap intervals with forecasts fixed. (d--f) Change in the $H_{40}$ increment after removing each additional predictor and refitting, $C_j$ (Equation~\ref{eq:heat_overlap}). Red indicates a larger heat-flux increment after removal; blue indicates a smaller increment. All comparisons use matched held-out pairs and retain the same 60-day $S_1$ history.}
\label{fig:heatflux}
\end{figure}

We next examine how these predictive contributions relate to accumulated lower-stratospheric eddy heat-flux anomalies using $H_{40}$ (Figures~\ref{fig:heatflux}a--c). At a 10-day lead for All winters, adding $H_{40}$ to the baseline reduces $Q$ by 0.014 (95\% interval 0.003--0.025), whereas adding the 12 eigen-microstate coefficients reduces it by 0.073 (0.048--0.098). Once those coefficients are included, the remaining $H_{40}$ increment is 0.004 ($-0.001$--0.009; Figure~\ref{fig:heatflux}a). In SSW winters the corresponding increments are 0.022, 0.080, and 0.005 (Figure~\ref{fig:heatflux}b). The short-lead heat-flux increment is smaller in normal winters (Figure~\ref{fig:heatflux}c), and no uniform positive increment persists at long leads. These contrasts indicate overlap between the forecast information supplied by $H_{40}$ and by the eigen-microstate coefficients, while the larger full-model gain shows that the latter is not represented by $H_{40}$ alone.

Deleting individual predictors localizes this overlap (Figures~\ref{fig:heatflux}d--f). The largest positive $C_j$ occurs for $S_6$, reaching 0.0113 at 8~days for All winters (Figure~\ref{fig:heatflux}d) and 0.0187 at 10~days for SSW winters (Figure~\ref{fig:heatflux}e). Thus, omitting $S_6$ makes $H_{40}$ more useful for forecasting SPV even though the same $S_1$ history and the other 11 coefficients remain in the model. This pattern is consistent with $S_6$ containing information related to accumulated wave activity and the associated stratospheric state. The radial structure of $S_6$ (Figure~\ref{fig:statesS}g) provides a description of stratospheric variability beyond bulk intensity, offering a plausible connection between this spatial coordinate and the integrated effects of wave activity. Negative values of $C_j$ elsewhere indicate that deleting that coefficient decreases, rather than increases, the incremental benefit from $H_{40}$.

At 100~hPa, eddy heat flux is related to upward wave activity entering the stratosphere, whose amplitude also depends on the background circulation and wave propagation \citep{Baldwin2021,BirnerAlbers2017}. The shared predictive information can therefore reflect both incoming disturbances and the evolving stratospheric response. This connects the eigen-microstate representation to stratosphere--troposphere coupling without assigning its bidirectional interactions to a unique forcing pathway.

\FloatBarrier

\section{Conclusion}
In this study, we provide a state-resolved and quantitative framework for understanding the predictability of SPV based on EMT and mesoscopic Granger-causality analysis. By decomposing atmospheric circulation into eigen microstates, we identify how both internal stratospheric variability and tropospheric variability contribute to SPV predictability.

Our results show that SPV variability is characterized by a hierarchy of stratospheric eigen microstates, ranging from the leading state representing bulk vortex intensity to higher-order states representing zonal asymmetry and radial structure. In contrast, tropospheric eigen microstates exhibit more complex multipolar structures and carry complementary predictive information. We demonstrate that SPV predictability is multimodal and strongly lead-time dependent. Short-term predictability is primarily governed by the persistence of the leading stratospheric state, reflecting the intrinsic memory of the vortex. Over the following several weeks, higher-order stratospheric states and tropospheric states provide predictive information beyond bulk vortex intensity. Leave-one-winter-out forecasts increase the all-winter 30-day correlation, but the additional gains diminish at longer leads and are not sustained uniformly to 60~days.

Predictive gains are particularly pronounced in December--January. Around SSWs, additional spatial information improves parts of the pre-event forecasts, while the relatively high skill later in the event is largely shared with the history baseline.

The heat-flux analysis shows that preceding 100~hPa eddy heat flux provides predictive information that overlaps with information carried by the additional eigen microstates, particularly $S_6$, linking their predictive contributions to wave activity.

These results demonstrate that SPV predictability is multimodal and can be quantitatively characterized within a unified, state-resolved framework. This approach extends predictability analysis beyond traditional low-dimensional indices and provides a physically interpretable pathway for improving subseasonal-to-seasonal forecasts.

\section*{Data availability}
ERA5 reanalysis data used in this study are publicly available
from the European Centre for Medium-Range Weather Forecasts
(ECMWF) Climate Data Store
\citep{ERA5PressureLevels}.
The analysis code and derived figure data are publicly
available at
\url{https://github.com/RichY-YikO/spv-emt-reproduction}
and archived on Zenodo \citep{YangSPVCode}
(version~v1.0.0;
\url{https://doi.org/10.5281/zenodo.23524214}).

\section*{Competing interests}
The authors declare there are no conflicts of interest for this study.

\section*{Acknowledgments}
We thank the National Key Research and Development Program of China (Grant No.\ 2023YFE0109000) and the National Natural Science Foundation of China (Grants No.~12305044 and No.~12135003) for financial support.

\section*{Author contributions}
Yongwen Zhang conceived the study, developed the conceptual framework and supervised the research. Shuo Yang and Dan Zhao curated and processed the data, developed the analysis code, performed the analyses, prepared the figures and wrote the initial manuscript. Xiaosong Chen contributed substantially to the interpretation of the results and critically revised the manuscript. Tingting Xue and Chunhua Zeng contributed to discussions of the results and manuscript revision. All authors revised, reviewed and approved the final manuscript.

\bibliographystyle{plainnat}
\bibliography{references}

@article{Baldwin1999,
  author = {Baldwin, Mark P. and Dunkerton, Timothy J.},
  title = {{Propagation of the Arctic Oscillation from the stratosphere to the troposphere}},
  year = {1999},
  journal = {Journal of Geophysical Research: Atmospheres},
  volume = {104},
  number = {D24},
  pages = {30937--30946},
  doi = {10.1029/1999jd900445},
}

@article{Baldwin2001,
  author = {Baldwin, Mark P. and Dunkerton, Timothy J.},
  title = {{Stratospheric Harbingers of Anomalous Weather Regimes}},
  year = {2001},
  journal = {Science},
  volume = {294},
  number = {5542},
  pages = {581--584},
  doi = {10.1126/science.1063315},
}

@article{Baldwin2021,
  author = {Baldwin, Mark P. and Ayarzag{\"{u}}ena, Blanca and Birner, Thomas and Butchart, Neal and Butler, Amy H. and Charlton-Perez, Andrew J. and Domeisen, Daniela I. V. and Garfinkel, Chaim I. and Garny, Hella and Gerber, Edwin P. and Hegglin, Michaela I. and Langematz, Ulrike and Pedatella, Nicholas M.},
  title = {{Sudden Stratospheric Warmings}},
  year = {2021},
  journal = {Reviews of Geophysics},
  volume = {59},
  number = {1},
  pages = {e2020RG000708},
  doi = {10.1029/2020rg000708},
}

@article{BirnerAlbers2017,
  author = {Birner, Thomas and Albers, John R.},
  title = {{Sudden Stratospheric Warmings and Anomalous Upward Wave Activity Flux}},
  year = {2017},
  journal = {SOLA},
  volume = {13A},
  pages = {8--12},
  doi = {10.2151/sola.13a-002},
}

@article{Butler2015,
  author = {Butler, Amy H. and Seidel, Dian J. and Hardiman, Steven C. and Butchart, Neal and Birner, Thomas and Match, Aaron},
  title = {{Defining Sudden Stratospheric Warmings}},
  year = {2015},
  journal = {Bulletin of the American Meteorological Society},
  volume = {96},
  number = {11},
  pages = {1913--1928},
  doi = {10.1175/bams-d-13-00173.1},
}

@incollection{Butler2018,
  author = {Butler, Amy H. and Charlton-Perez, Andrew J. and Domeisen, Daniela I. V. and Garfinkel, Chaim I. and Gerber, Edwin P. and Hitchcock, Peter and Karpechko, Alexey Yu. and Maycock, Amanda C. and Sigmond, Michael and Simpson, Isla R. and Son, Seok-Woo},
  title = {{Sub-seasonal Predictability and the Stratosphere}},
  year = {2019},
  booktitle = {{Sub-Seasonal to Seasonal Prediction: The Gap Between Weather and Climate Forecasting}},
  editor = {Robertson, Andrew W. and Vitart, Fr{\'{e}}d{\'{e}}ric},
  publisher = {Elsevier},
  pages = {223--241},
  doi = {10.1016/b978-0-12-811714-9.00011-5},
}

@article{CameronMiller2015,
  author = {Cameron, A. Colin and Miller, Douglas L.},
  title = {{A Practitioner's Guide to Cluster-Robust Inference}},
  year = {2015},
  journal = {Journal of Human Resources},
  volume = {50},
  number = {2},
  pages = {317--372},
  doi = {10.3368/jhr.50.2.317},
}

@article{Charlton2007,
  author = {Charlton, Andrew J. and Polvani, Lorenzo M.},
  title = {{A New Look at Stratospheric Sudden Warmings. Part I: Climatology and Modeling Benchmarks}},
  year = {2007},
  journal = {Journal of Climate},
  volume = {20},
  number = {3},
  pages = {449--469},
  doi = {10.1175/jcli3996.1},
}

@article{Charney1961,
  author = {Charney, J. G. and Drazin, P. G.},
  title = {{Propagation of planetary-scale disturbances from the lower into the upper atmosphere}},
  year = {1961},
  journal = {Journal of Geophysical Research},
  volume = {66},
  number = {1},
  pages = {83--109},
  doi = {10.1029/jz066i001p00083},
}

@article{ChenWei2009,
  author = {Chen, Wen and Wei, Ke},
  title = {{Interannual variability of the winter stratospheric polar vortex in the Northern Hemisphere and their relations to QBO and ENSO}},
  year = {2009},
  journal = {Advances in Atmospheric Sciences},
  volume = {26},
  number = {5},
  pages = {855--863},
  doi = {10.1007/s00376-009-8168-6},
}

@article{Domeisen2020a,
  author = {Domeisen, Daniela I. V. and Butler, Amy H. and Charlton-Perez, Andrew J. and Ayarzag{\"{u}}ena, Blanca and Baldwin, Mark P. and Dunn-Sigouin, Etienne and Furtado, Jason C. and Garfinkel, Chaim I. and Hitchcock, Peter and Karpechko, Alexey Yu. and Kim, Hera and Knight, Jeff and Lang, Andrea L. and Lim, Eun-Pa and Marshall, Andrew and Roff, Greg and Schwartz, Chen and Simpson, Isla R. and Son, Seok-Woo and Taguchi, Masakazu},
  title = {{The Role of the Stratosphere in Subseasonal to Seasonal Prediction: 1. Predictability of the Stratosphere}},
  year = {2020},
  journal = {Journal of Geophysical Research: Atmospheres},
  volume = {125},
  number = {2},
  pages = {e2019JD030920},
  doi = {10.1029/2019jd030920},
}

@article{Domeisen2020b,
  author = {Domeisen, Daniela I. V. and Butler, Amy H. and Charlton-Perez, Andrew J. and Ayarzag{\"{u}}ena, Blanca and Baldwin, Mark P. and Dunn-Sigouin, Etienne and Furtado, Jason C. and Garfinkel, Chaim I. and Hitchcock, Peter and Karpechko, Alexey Yu. and Kim, Hera and Knight, Jeff and Lang, Andrea L. and Lim, Eun-Pa and Marshall, Andrew and Roff, Greg and Schwartz, Chen and Simpson, Isla R. and Son, Seok-Woo and Taguchi, Masakazu},
  title = {{The Role of the Stratosphere in Subseasonal to Seasonal Prediction: 2. Predictability Arising From Stratosphere-Troposphere Coupling}},
  year = {2020},
  journal = {Journal of Geophysical Research: Atmospheres},
  volume = {125},
  number = {2},
  pages = {e2019JD030923},
  doi = {10.1029/2019jd030923},
}

@misc{ERA5PressureLevels,
  author = {{Copernicus Climate Change Service}},
  title = {{ERA5 hourly data on pressure levels from 1940 to present}},
  year = {2018},
  howpublished = {[Data set]. Copernicus Climate Change Service (C3S) Climate Data Store (CDS)},
  doi = {10.24381/cds.bd0915c6},
  url = {https://cds.climate.copernicus.eu/datasets/reanalysis-era5-pressure-levels},
}

@article{Granger1969,
  author = {Granger, C. W. J.},
  title = {{Investigating Causal Relations by Econometric Models and Cross-spectral Methods}},
  year = {1969},
  journal = {Econometrica},
  volume = {37},
  number = {3},
  pages = {424--438},
  doi = {10.2307/1912791},
}

@article{Gray2020,
  author = {Gray, L. J. and Brown, M. J. and Knight, J. and Andrews, M. and Lu, H. and O'Reilly, C. and Anstey, J.},
  title = {{Forecasting extreme stratospheric polar vortex events}},
  year = {2020},
  journal = {Nature Communications},
  volume = {11},
  number = {1},
  pages = {4630},
  doi = {10.1038/s41467-020-18299-7},
}

@article{Hannachi2007,
  author = {Hannachi, A. and Jolliffe, I. T. and Stephenson, D. B.},
  title = {{Empirical orthogonal functions and related techniques in atmospheric science: A review}},
  year = {2007},
  journal = {International Journal of Climatology},
  volume = {27},
  number = {9},
  pages = {1119--1152},
  doi = {10.1002/joc.1499},
}

@article{Hersbach2020,
  author = {Hersbach, Hans and Bell, Bill and Berrisford, Paul and Hirahara, Shoji and Hor{\'{a}}nyi, Andr{\'{a}}s and Mu{\~{n}}oz-Sabater, Joaqu{\'{i}}n and Nicolas, Julien and Peubey, Carole and Radu, Raluca and Schepers, Dinand and Simmons, Adrian and Soci, Cornel and Abdalla, Saleh and Abellan, Xavier and Balsamo, Gianpaolo and Bechtold, Peter and Biavati, Gionata and Bidlot, Jean and Bonavita, Massimo and De Chiara, Giovanna and Dahlgren, Per and Dee, Dick and Diamantakis, Michail and Dragani, Rossana and Flemming, Johannes and Forbes, Richard and Fuentes, Manuel and Geer, Alan and Haimberger, Leo and Healy, Sean and Hogan, Robin J. and H{\'{o}}lm, El{\'{i}}as and Janiskov{\'{a}}, Marta and Keeley, Sarah and Laloyaux, Patrick and Lopez, Philippe and Lupu, Cristina and Radnoti, Gabor and de Rosnay, Patricia and Rozum, Iryna and Vamborg, Freja and Villaume, Sebastien and Th{\'{e}}paut, Jean-No{\"{e}}l},
  title = {{The ERA5 global reanalysis}},
  year = {2020},
  journal = {Quarterly Journal of the Royal Meteorological Society},
  volume = {146},
  number = {730},
  pages = {1999--2049},
  doi = {10.1002/qj.3803},
}

@article{Holton1976,
  author = {Holton, James R. and Mass, Clifford},
  title = {{Stratospheric Vacillation Cycles}},
  year = {1976},
  journal = {Journal of the Atmospheric Sciences},
  volume = {33},
  number = {11},
  pages = {2218--2225},
  doi = {10.1175/1520-0469(1976)033<2218:svc>2.0.co;2},
}

@article{Hu2019,
  author = {Hu, Gaoke and Liu, Teng and Liu, Maoxin and Chen, Wei and Chen, Xiaosong},
  title = {{Condensation of eigen microstate in statistical ensemble and phase transition}},
  year = {2019},
  journal = {Science China Physics, Mechanics \& Astronomy},
  volume = {62},
  number = {9},
  pages = {990511},
  doi = {10.1007/s11433-018-9353-x},
}

@book{Jolliffe2002,
  author = {Jolliffe, I. T.},
  title = {{Principal Component Analysis}},
  year = {2002},
  publisher = {Springer-Verlag},
  address = {New York},
  series = {Springer Series in Statistics},
  edition = {2},
  doi = {10.1007/b98835},
}

@article{Kidston2015,
  author = {Kidston, Joseph and Scaife, A. A. and Hardiman, Steven C. and Mitchell, Daniel M. and Butchart, Neal and Baldwin, Mark P. and Gray, L. J.},
  title = {{Stratospheric influence on tropospheric jet streams, storm tracks and surface weather}},
  year = {2015},
  journal = {Nature Geoscience},
  volume = {8},
  number = {6},
  pages = {433--440},
  doi = {10.1038/ngeo2424},
}

@article{Kretschmer2017,
  author = {Kretschmer, Marlene and Runge, Jakob and Coumou, Dim},
  title = {{Early prediction of extreme stratospheric polar vortex states based on causal precursors}},
  year = {2017},
  journal = {Geophysical Research Letters},
  volume = {44},
  number = {16},
  pages = {8592--8600},
  doi = {10.1002/2017gl074696},
}

@article{Ma2024,
  author = {Ma, Xuan and Liang, Rizhou and Chen, Xiaosong and Xie, Fei and Zuo, Jinqing and Sun, Cheng and Ding, Ruiqiang},
  title = {{Increased Predictability of Extreme El Ni{\~{n}}o From Decadal Interbasin Interaction}},
  year = {2024},
  journal = {Geophysical Research Letters},
  volume = {51},
  number = {23},
  pages = {e2024GL110943},
  doi = {10.1029/2024gl110943},
}

@article{Matsuno1971,
  author = {Matsuno, Taroh},
  title = {{A Dynamical Model of the Stratospheric Sudden Warming}},
  year = {1971},
  journal = {Journal of the Atmospheric Sciences},
  volume = {28},
  number = {8},
  pages = {1479--1494},
  doi = {10.1175/1520-0469(1971)028<1479:admots>2.0.co;2},
}

@article{MatthewmanEsler2011,
  author = {Matthewman, N. Joss and Esler, J. G.},
  title = {{Stratospheric Sudden Warmings as Self-Tuning Resonances. Part I: Vortex Splitting Events}},
  year = {2011},
  journal = {Journal of the Atmospheric Sciences},
  volume = {68},
  number = {11},
  pages = {2481--2504},
  doi = {10.1175/jas-d-11-07.1},
}

@article{Monahan2009,
  author = {Monahan, Adam H. and Fyfe, John C. and Ambaum, Maarten H. P. and Stephenson, D. B. and North, Gerald R.},
  title = {{Empirical Orthogonal Functions: The Medium is the Message}},
  year = {2009},
  journal = {Journal of Climate},
  volume = {22},
  number = {24},
  pages = {6501--6514},
  doi = {10.1175/2009jcli3062.1},
}

@article{PolvaniWaugh2004,
  author = {Polvani, Lorenzo M. and Waugh, Darryn W.},
  title = {{Upward Wave Activity Flux as a Precursor to Extreme Stratospheric Events and Subsequent Anomalous Surface Weather Regimes}},
  year = {2004},
  journal = {Journal of Climate},
  volume = {17},
  number = {18},
  pages = {3548--3554},
  doi = {10.1175/1520-0442(2004)017<3548:uwafaa>2.0.co;2},
}

@article{Quadrelli2004,
  author = {Quadrelli, Roberta and Wallace, John M.},
  title = {{A Simplified Linear Framework for Interpreting Patterns of Northern Hemisphere Wintertime Climate Variability}},
  year = {2004},
  journal = {Journal of Climate},
  volume = {17},
  number = {19},
  pages = {3728--3744},
  doi = {10.1175/1520-0442(2004)017<3728:aslffi>2.0.co;2},
}

@article{Richter2020,
  author = {Richter, Jadwiga H. and Pegion, Kathy and Sun, Lantao and Kim, Hyemi and Caron, Julie M. and Glanville, Anne and LaJoie, Emerson and Yeager, Stephen and Kim, Who M. and Tawfik, Ahmed and Collins, Dan},
  title = {{Subseasonal Prediction with and without a Well-Represented Stratosphere in CESM1}},
  year = {2020},
  journal = {Weather and Forecasting},
  volume = {35},
  number = {6},
  pages = {2589--2602},
  doi = {10.1175/waf-d-20-0029.1},
}

@article{Runge2019,
  author = {Runge, Jakob and Bathiany, Sebastian and Bollt, Erik and Camps-Valls, Gustau and Coumou, Dim and Deyle, Ethan and Glymour, Clark and Kretschmer, Marlene and Mahecha, Miguel D. and Mu{\~{n}}oz-Mar{\'{i}}, Jordi and van Nes, Egbert H. and Peters, Jonas and Quax, Rick and Reichstein, Markus and Scheffer, Marten and Sch{\"{o}}lkopf, Bernhard and Spirtes, Peter and Sugihara, George and Sun, Jie and Zhang, Kun and Zscheischler, Jakob},
  title = {{Inferring causation from time series in Earth system sciences}},
  year = {2019},
  journal = {Nature Communications},
  volume = {10},
  number = {1},
  pages = {2553},
  doi = {10.1038/s41467-019-10105-3},
}

@article{Scaife2014,
  author = {Scaife, A. A. and Arribas, A. and Blockley, E. and Brookshaw, A. and Clark, R. T. and Dunstone, N. and Eade, R. and Fereday, D. and Folland, C. K. and Gordon, M. and Hermanson, L. and Knight, J. R. and Lea, D. J. and MacLachlan, C. and Maidens, A. and Martin, M. and Peterson, A. K. and Smith, D. and Vellinga, M. and Wallace, E. and Waters, J. and Williams, A.},
  title = {{Skillful long-range prediction of European and North American winters}},
  year = {2014},
  journal = {Geophysical Research Letters},
  volume = {41},
  number = {7},
  pages = {2514--2519},
  doi = {10.1002/2014gl059637},
}

@article{Shan2022,
  author = {Shan, Qi and Fan, Ke},
  title = {{The Transition of Stratospheric Polar Vortex Intensity: A Case Study of Winter 1987/88}},
  year = {2022},
  journal = {Journal of Geophysical Research: Atmospheres},
  volume = {127},
  number = {12},
  pages = {e2022JD036511},
  doi = {10.1029/2022jd036511},
}

@article{Sun2021,
  author = {Sun, Yu and Hu, Gaoke and Zhang, Yongwen and Lu, Bo and Lu, Zhenghui and Fan, Jingfang and Li, Xiaoteng and Deng, Qimin and Chen, Xiaosong},
  title = {{Eigen microstates and their evolutions in complex systems}},
  year = {2021},
  journal = {Communications in Theoretical Physics},
  volume = {73},
  number = {6},
  pages = {065603},
  doi = {10.1088/1572-9494/abf127},
}

@article{Thompson2000,
  author = {Thompson, David W. J. and Wallace, John M.},
  title = {{Annular Modes in the Extratropical Circulation. Part I: Month-to-Month Variability}},
  year = {2000},
  journal = {Journal of Climate},
  volume = {13},
  number = {5},
  pages = {1000--1016},
  doi = {10.1175/1520-0442(2000)013<1000:amitec>2.0.co;2},
}

@article{Thompson2000b,
  author = {Thompson, David W. J. and Wallace, John M. and Hegerl, Gabriele C.},
  title = {{Annular Modes in the Extratropical Circulation. Part II: Trends}},
  year = {2000},
  journal = {Journal of Climate},
  volume = {13},
  number = {5},
  pages = {1018--1036},
  doi = {10.1175/1520-0442(2000)013<1018:amitec>2.0.co;2},
}

@article{Tripathi2015,
  author = {Tripathi, Om P and Charlton-Perez, Andrew J. and Sigmond, Michael and Vitart, Frederic},
  title = {{Enhanced long-range forecast skill in boreal winter following stratospheric strong vortex conditions}},
  year = {2015},
  journal = {Environmental Research Letters},
  volume = {10},
  number = {10},
  pages = {104007},
  doi = {10.1088/1748-9326/10/10/104007},
}

@article{WuRW2022,
  author = {Wu, Rachel Wai-Ying and Wu, Zheng and Domeisen, Daniela I. V.},
  title = {{Differences in the sub-seasonal predictability of extreme stratospheric events}},
  year = {2022},
  journal = {Weather and Climate Dynamics},
  volume = {3},
  number = {3},
  pages = {755--776},
  doi = {10.5194/wcd-3-755-2022},
}

@article{WuZ2022,
  author = {Wu, Zheng and Beucler, Tom and Sz{\'{e}}kely, Enik{\H{o}} and Ball, William T. and Domeisen, Daniela I. V.},
  title = {{Modeling stratospheric polar vortex variation and identifying vortex extremes using explainable machine learning}},
  year = {2022},
  journal = {Environmental Data Science},
  volume = {1},
  pages = {e17},
  doi = {10.1017/eds.2022.19},
}

@article{Zhang2024,
  author = {Zhang, Yongwen and Liu, Maoxin and Hu, Gaoke and Liu, Teng and Chen, Xiaosong},
  title = {{Eigen microstates in self-organized criticality}},
  year = {2024},
  journal = {Physical Review E},
  volume = {109},
  number = {4},
  pages = {044130},
  doi = {10.1103/physreve.109.044130},
}

@misc{YangSPVCode,
  author    = {Yang, Shuo},
  year      = {2026},
  title     = {Code and reproduction materials for
               ``State-resolved multimodal contributions to
               stratospheric polar vortex predictability''},
  version   = {v1.0.0},
  publisher = {Zenodo},
  doi       = {10.5281/zenodo.23524214},
  url       = {https://doi.org/10.5281/zenodo.23524214},
  note      = {[Software]}
}
\end{document}